\documentclass[11pt,twoside]{article}

\usepackage{asp2006}
\usepackage{epsf}
\usepackage{lscape}
\usepackage{graphicx}

\begin{document}
\title{Evolution of Star-forming Galaxies seen from Lyman Break Galaxies
at $z\sim5$}   
\author{K. Ohta\altaffilmark{1}, K. Yabe\altaffilmark{1}, 
I. Iwata\altaffilmark{2}, M. Sawicki\altaffilmark{3},
M. Akiyama\altaffilmark{4}, N. Tamura\altaffilmark{4}, 
and K. Aoki\altaffilmark{4}}
\altaffiltext{1}{Department of Astronomy, Kyoto University, Japan}
\altaffiltext{2}{Okayama Astrophysical Observatory, National Astronomical
Observatory, Japan}
\altaffiltext{3}{Department of Astronomy and Physics, St. Mary's University,
Canada}
\altaffiltext{4}{Subaru Telescope, National Astronomical Observatory of Japan, USA}

\begin{abstract} 
Properties of Lyman Break Galaxies at $z\sim5$ are presented and
are compared with those at $z\sim 2-3$ LBGs to see evolution of LBGs
from $z\sim5$ to $z \sim3-2$.
The LBGs at $z\sim5$ are tend to have smaller stellar masses, but they
may show large star formation rates and specific star formation rates.  
\end{abstract}



\section{Introduction}

We have been studying properties of Lyman Break Galaxies (LBGs) at $z\sim5$,
$\sim 1$ Gyr after the Big Bang and before $z\sim3$.
In this contribution, we present our recent results obtained for $z\sim5$ LBGs
and possible evolution from $z\sim5$ LBGs  to $z \sim 2-3$ LBGs.
Our sample at $z\sim5$ was made in a field including the Hubble Deep Field 
North (HDF-N)
 by using the Subaru SCam through $V$-, $I_C$-, and $z'$-filters.
Since in and around the HDF-N field many spectroscopic data were available,
the color criteria to select the $z\sim 5$ LBGs can be reliably examined.
A sample of $z\sim5$ LBGs was chosen from galaxies brighter than
$z^{\prime}=26.5$ mag.
A number of the sample is $\sim 600$ in an area of $\sim500$ arcmin$^2$.

\section{Luminosity dependent evolution?}
UV luminosity function is presented by Iwata et al. (2003, 2007, 
2008 this volume).
In their results, from $z\sim 5$ to $\sim3$, 
a number of faint LBGs (fainter than $L_*$) is increasing with redshift, 
while in the brighter part the number does not change significantly.
Although there is a controversy on this result, a luminosity dependent 
evolution of UV luminosity function is indicated.

Spectroscopic follow-up observations of a part of our LBG
sample were made by Ando et al. (2004).
They found that UV luminous LBGs (brighter than $L_*$) do not
show large equivalent width of Ly$\alpha$ emission line.
Ando et al. (2006) showed this trend together with other LBGs as well as
Ly$\alpha$ emitters (LAEs).
This may be due to the dusty environment 
or the presence of more massive HI gas in the luminous LBGs.

\section{Stellar masses and star formation rates in $z\sim5$ LBGs}
Using public IRAC data in the GOODS-N region and those obtained in the
flanking fields, we constructed a  $z\sim5$ LBG sample with IRAC photometry.
From the sample of the LBGs mentioned above, we selected the LBGs those are
free from the contamination by neighboring objects by eye inspection.
The sample size is $\sim 170$; about half of them are detected  both in
IRAC channels 1 and 2.
We derived the Spectral Energy Distributions (SEDs) for the sample,
and made SED fitting.
The spectral model adopted for the SED fitting we use is rather traditional, because we
intend to compare with the results obtained in the previous studies 
at $z\sim 2-3$;
stellar synthesis code by Bruzual and Charlot (2003) with 0.2 $Z_{\odot}$,
the Salpeter IMF ($0.1- 100 M_{\odot}$) under
constant star formation history, and Calzetti extinction curve
(Calzetti et al. 2000).
We include the H$\alpha$ emission line calculated from the model intrinsic star
formation rate and extinction.
The SED fitting method is the essentially the same as that by
Sawicki and Yee (1998).  
In the fitting, we fixed the redshift to be 4.8.

The resulting stellar masses, star formation ages, color excess, and
star formation rates are presented in Figure \ref{z5lbg} (left)
overlayed with
those obtained for $z\sim3$ LBGs (Shapley et al. 2001) and for $z\sim 2$
LBGs (Shapley et al. 2005). 
The rest-frame UV luminosity and optical luminosity of these samples are
mostly overlapped each other.
The stellar masses of the $z\sim5$ LBGs are significantly smaller than those
at $z\sim 2-3$, and  ages are smaller on average.
These suggest the premature nature of the $z\sim 5$ LBGs.
Stellar mass function and stellar mass density at $z\sim5$ are
presented by Yabe et al. (2008, this volume).

\section{LBGs with large SFRs?}
\begin{figure}
\includegraphics[angle=-90,scale=0.4]{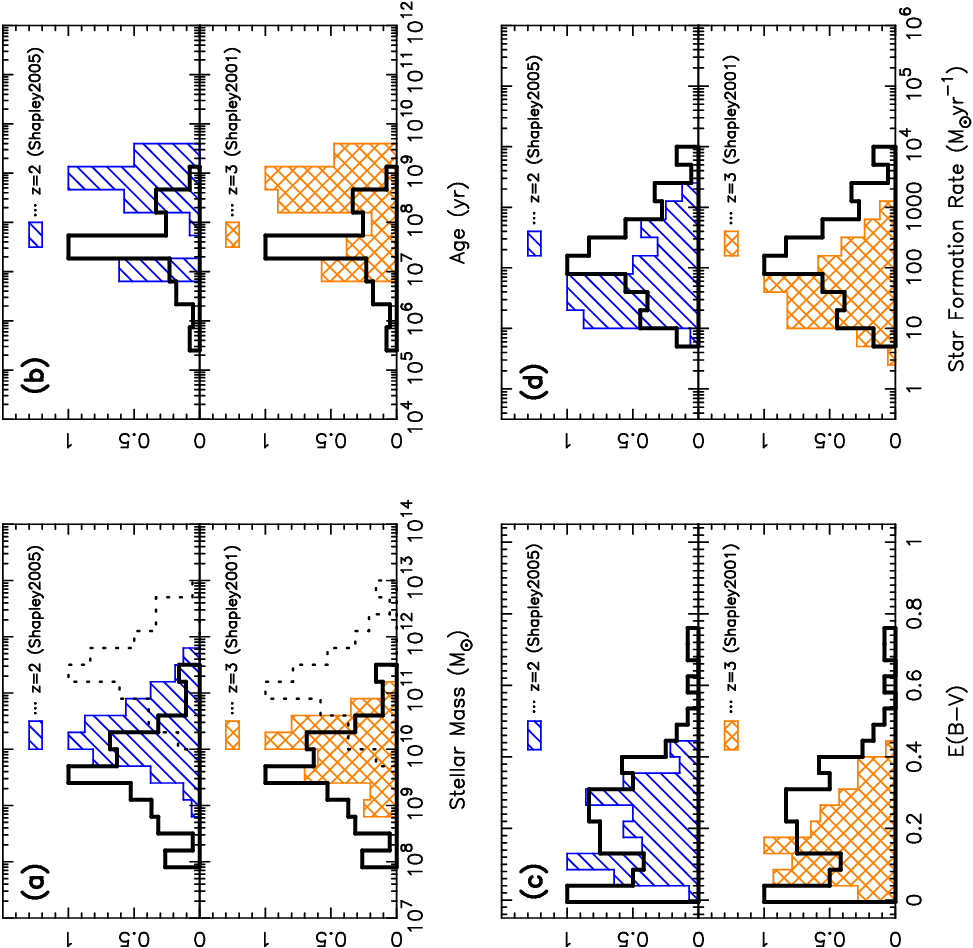}
\includegraphics[angle=-90,scale=0.4]{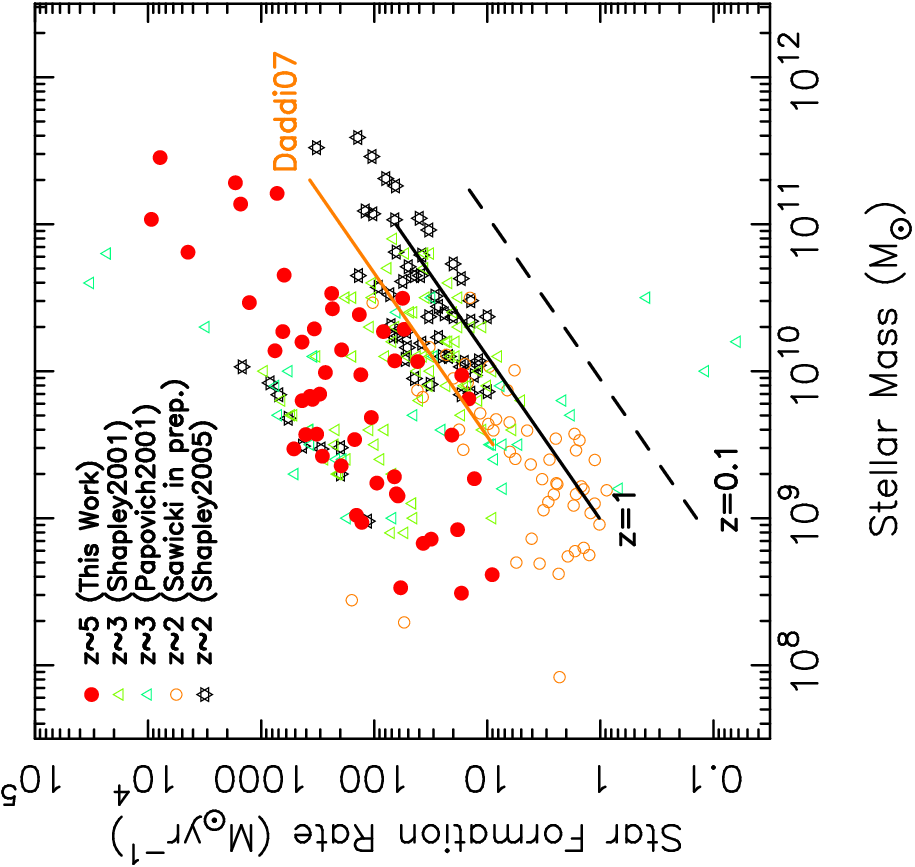}
\caption{{\bf Left:} Stellar mass (upper left), age (upper right), 
color excess (lower left), and star formation rate (lower right) 
of the SED fitting are shown with solid lines. Also shown are
those for $z\sim2$ and $z\sim3$ LBGs (Shapley et al. 2005; 
Shapley et al. 2001).
{\bf Right:} Star formation rate vs stellar mass for $z\sim 5$ LBGs (filled
circles).  Also shown are those for $z\sim 2-3$ LBGs (open symbols).
Solid lines show regression lines for star forming galaxies at
the redshifts shown taken from Daddi et al. (2007).}\label{z5lbg}
\end{figure}

As seen from Figure \ref{z5lbg} (left), the star formation rates (SFRs)
are systematically larger than those in $z\sim 2-3$ LBGs.
Figure \ref{z5lbg} (right) shows SFRs against their 
stellar masses for $z\sim5$ LBGs (filled circles).
The SFRs are very high amounting to several 
hundreds $M_{\odot}$ yr$^{-1}$.
These values are significantly larger than those seen in $z\sim 2-3$ LBGs
shown as open symbols in Figure \ref{z5lbg} (right).
They are also larger compared with the results for $z \sim 0.1-2$
star forming galaxies (e.g., Daddi et al. 2007).
The specific star formation rates tend to decrease with decreasing redshift.
It is suggested that the $z\sim 5$  LBGs are in active star forming phase
and have dusty environment. 
In fact, the color excesses obtained for the $z\sim5$ LBGs tend to be larger 
than those obtained for $z\sim2-3$ LBGs (Figure \ref{z5lbg}(left)).
We may be witnessing an emergence  of a population of active star forming
galaxies at $z\sim5$.

If the $z\sim5$ LBGs keep the high SFRs, they should have stellar masses
 more than $10^{11} M_{\odot}$ at $z\sim3$.
The distributions of the expected stellar masses at $z\sim 2 -3$ are shown 
as dotted lines in Figure \ref{z5lbg}(left), suggesting that the
$z\sim 5$ LBGs are not the progenitor of the $z\sim2-3$ LBGs or
the large SFRs should decline.
If the SFRs decline exponentially with a time scale of $10^8$ yr, 
SFRs should be very much small at $z\sim3$.
Thus the star formation may be episodic if the $z\sim5$ LBGs are the 
progenitor of $z\sim3$ LBGs.
Or they might be  progenitor of passive massive galaxies at $z=2-3$.

\begin{figure}
\includegraphics[angle=-90,scale=0.4]{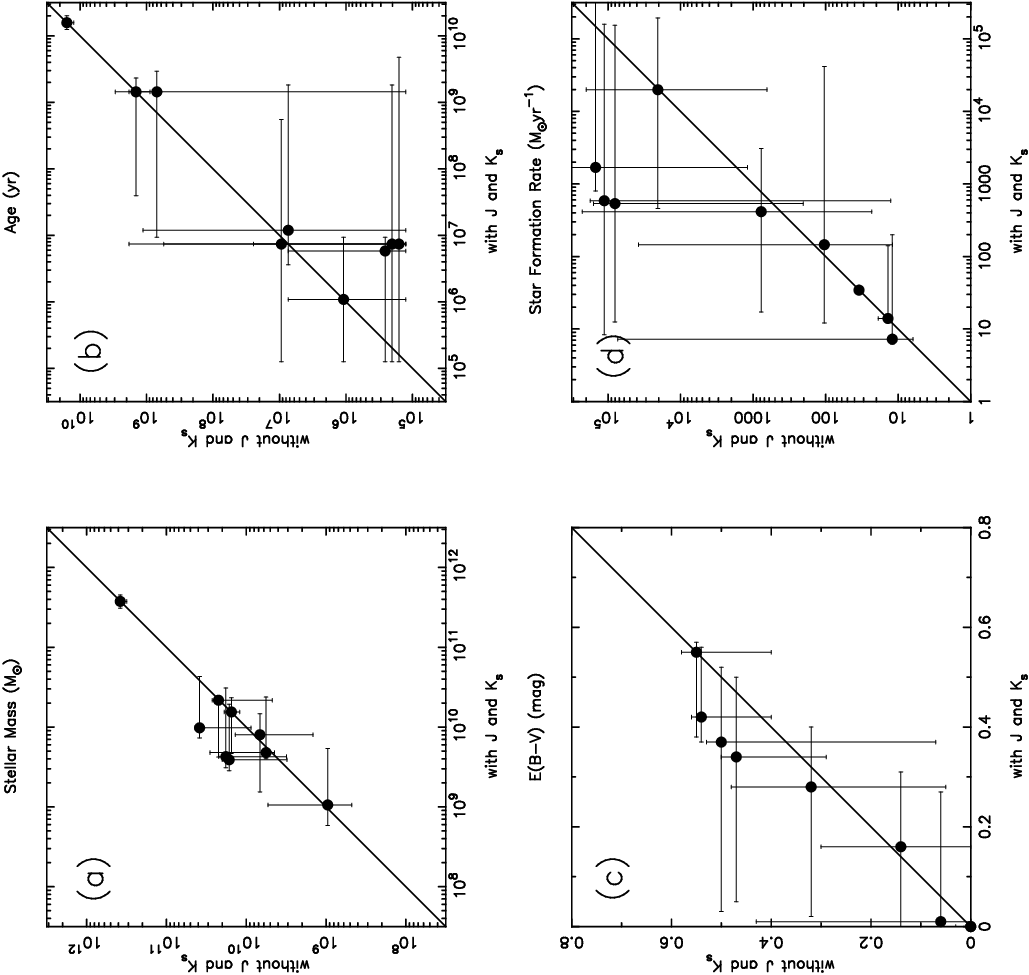}
\includegraphics[angle=-90,scale=0.4]{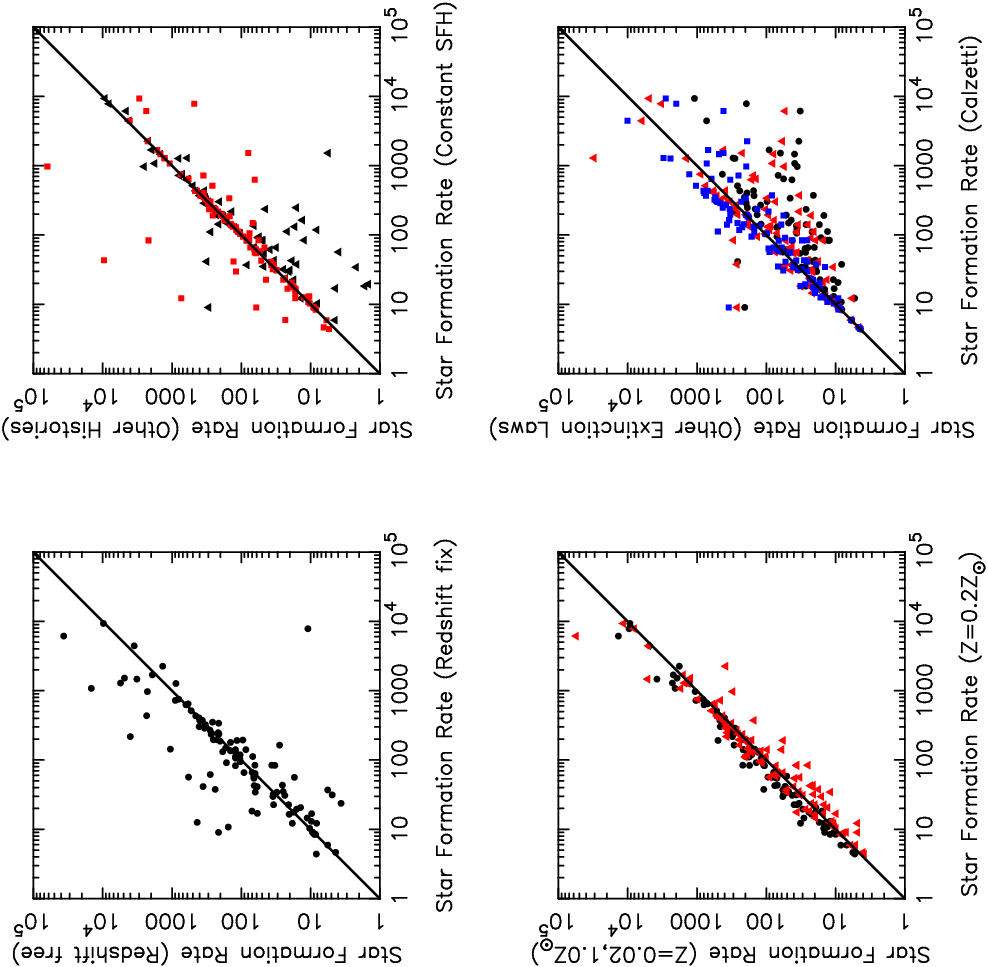}
\caption{Uncertainty  of SED fitting.
{\bf Left:} stellar mass, age, color excess, and star formation rate
obtained from the SEDs by Stark et al. (2007)
with (abscissas) and without (ordinates) $J$ and $K$ data.
{\bf Right:}  Star formation rates using fiducial model (abscissas) and
other models (ordinates).
Cases by adopting free parameter redshift, star formation history of 
exponentially decaying model (triangles) and two-component model (squares),
metallicity of 1.0 $Z_{\odot}$ (circles)
 and 0.1 $Z_{\odot}$ (triangles), and 
extinction curves of the Milky Way (squares), the LMC
(triangles), and the SMC (circles) are shown.
 }\label{stark}
\end{figure}

However, as known well, there may be an age-extinction degeneracy in the SED
fitting.
Because we do not have NIR photometric data, our SED fitting may suffer from
the degeneracy.
To examine this, we made SED fitting for LBGs at $z\sim 5$ sampled by
Stark et al. (2007) in which $J$ and $K$ photometry are available.
We made SED fitting with and without $J$ and $K$ data.
The resulting output parameters are shown in Figure \ref{stark} (left).
The stellar masses are robust as seen from the figure.
The SFRs are also rather robust (except for a few extraordinary cases), 
though the error bars are large.
We also estimated  the color excess using the method by Meurer et al. (1995),
and found many of the objects have the color excesses consistent with the
SED fitting.
We also made various tests to see whether the large SFRs are due to some 
artificial causes.
Figure \ref{stark} (right) shows output parameters of the SED fitting
when we take redshift as a free parameter (upper left panel), other star
formation history (upper right panel), metallicity (lower left panel),
and other extinction curves (lower right panel). 
As seen from the panels, except for the adoption of other extinction curves,
majority of estimated SFRs agree with those obtained in the fiducial model.

No LBGs with large SFRs coincide
with SCUBA sources (Pope et al. 2005), X-ray sources (Alexander et al. 2003),
 and VLA sources (Richards 2000).
The estimated upper limits on SFRs from these surveys are $\sim 1500 M_{\odot}$
yr$^{-1}$.  
It is worth noting that many of these LBGs with large
SFRs do not reside in the SCUBA/Chandra surveys.
Further observations are necessary to definitively conclude
the presence of such population.

\begin{thebibliography}{}
\bibitem[Alexander et al. 2003]{Alex03} Alexander, D. M., et al. 2003, AJ, 126, 539
\bibitem[Ando et al. 2004]{Ando04} Ando, M., et al. 2004, ApJ, 610, 635
\bibitem[Ando et al. 2006]{Ando06} Ando, M., et al. 2006, ApJ, 645, L9
\bibitem[Bruzual and Charlot 2003]{BC03} Bruzual, A. G., \& Charlot, S. 2003, MNRAS, 344, 1000
\bibitem[Calzetti et al. 2000]{Cal00} Calzetti, D., et al. 2000, ApJ, 533, 682
\bibitem[Daddi et al. 2007]{Dad07} Daddi, E., et al. 2007, ApJ, 670, 156
\bibitem[Iwata et al. 2003]{Iwata03} Iwata, I., et al. 2003, PASJ, 55, 415
\bibitem[Iwata et al. 2007]{Iwata07} Iwata, I., et al. 2007, MNRAS, 376, 1557
\bibitem[Meurer et al. 1995]{Meu95} Meurer, G. R., Heckman, T. M., \& Calzetti, D. 1995, ApJ 521, 64
\bibitem[Pope et al. 2005]{Pope05} Pope, A., et al. 2005, MNRAS, 358, 149
\bibitem[Richards 2000]{Rich00} Richards, E. A. 2000, ApJ, 533, 611
\bibitem[Sawicki and Yee 1998]{Saw98} Sawicki, M., \& Yee, H. K. C. 1998, AJ, 115, 1329
\bibitem[Shaplaey et al. 2001]{Shap00} Shapley, A. E., et al. 2001, ApJ, 562, 95
\bibitem[Shapley et al. 2005]{Shap05} Shapley, A. E., et al. 2005, ApJ, 626, 698
\bibitem[Stark et al. 2007]{Stark07} Stark, D. P., et al. 2007, ApJ, 659, 84
\end{thebibliography}
\end{document}